\documentclass[8pt]{article}

\usepackage{amsmath,amssymb,amsthm,graphicx,latexsym}

\newcommand{\ed}{\end{document}}
\newcommand{\beq}{\begin{equation}}
\newcommand{\eeq}{\end{equation}}

\begin{document}
\begin{center}
\large{\textbf{\textbf{Path Integral for non-relativistic Generalized Uncertainty Principle corrected Hamiltonian}}}\\
\end{center}
\begin{center}
Sudipta Das\footnote{E-mail: sudipta.das\_r@isical.ac.in} and
Souvik Pramanik\footnote{E-mail: souvick.in@gmail.com} \\
Physics and Applied Mathematics Unit, Indian Statistical
Institute\\
203 B. T. Road, Kolkata 700108, India \\
\end{center}\vspace{0.5cm}

\begin{center}
{\textbf{Abstract}}
\end{center}
\emph{Generalized Uncertainty Principle (GUP) has brought the idea of existence of minimum measurable length
in Quantum physics. Depending on this GUP, non-relativistic Hamiltonian at the Planck scale is modified.
In this article, we construct the kernel for this GUP corrected Hamiltonian for free particle by applying
the Hamiltonian path integral approach and check the validity conditions for this kernel thoroughly.
Interestingly, the probabilistic interpretation of this kernel induces a momentum upper bound in the
theory which is comparable with GUP induced maximum momentum uncertainty.}\\

\vskip .3cm

\section{Introduction:}
There are many indications that in quantum gravity there might exist a minimal
observable distance of the order of the Planck length, $l_{pl}\approx~10^{-33}~cm$.
Generalized Uncertainty Principle (GUP) \cite{ven} naturally encodes the idea of
existence of a minimum measurable length through modifications in the Poisson
brackets of position $x$ and momentum $p$.
The Heisenberg Uncertainty Principle (HUP) says that uncertainty in position decreases with increasing
energy $(\Delta x\sim\frac{\hbar}{\Delta p})$. But HUP breaks down for energies
close to Planck scale, at which point the Schwarzschild radius becomes comparable
to Compton wavelength. Higher energies result in a further increase
of the Schwarzschild radius, inducing the following relation:
$\Delta x\approx l_{pl}^{2}\frac{\Delta p}{\hbar}$.
Consistent with the above, the following form of GUP has been proposed, postulated to hold
in all scales \cite{kempf}
\beq\Delta x_{i}\Delta p_{i}~\geq~\frac{\hbar}{2}\left[1+\beta(\Delta p^{2}+\langle
p\rangle^{2})+2\beta(\Delta p_{i}^{2}+<p_{i}>^{2})\right]~~~~i=1,2,3 \label{deltaxi} \eeq
where $[\beta]=(momentum)^{-2}$ and we will assume that
$\beta=\beta_{0}/(M_{pl}c)^{2}=l_{pl}^{2}/2\hbar^{2}$, $M_{pl}=$ Planck mass, and $M_{pl}c^{2}=$
Planck energy $\approx~10^{19}~GeV$. It is evident that the parameter $\beta_{0}$ is dimensionless and
normally assumed to be $\beta_{0}\approx~1$. In one dimension the above inequality takes the form:
\beq \Delta x\Delta p~\geq~\frac{\hbar}{2}\left[1+3\beta(\Delta p^{2}+\langle p\rangle^{2})\right] \label{delta1} \eeq
from which we have
\beq \Delta p\leq \frac{\Delta x}{3\beta\hbar}+\sqrt{\left(\frac{\Delta x}{3\beta\hbar}
\right)^{2}-\frac{1+3\beta\langle p\rangle^{2}}{3\beta}}. \label{deltafrac} \eeq
Since $\Delta p$ is real quantity, we have
\beq \left(\frac{\Delta x}{3\beta\hbar}\right)^{2}\geq\frac{1+3\beta\langle p\rangle^{2}}{3\beta}
\Rightarrow \Delta x \geq\hbar\sqrt{3\beta}\sqrt{1+3\beta\langle p\rangle^{2}} \label{deltafrac2} \eeq
which gives the minimum bound for $\Delta x$ as:
\beq
\Delta x_{min}=\hbar\sqrt{3\beta}. \label{xmin}
\eeq
Here one should notice that the condition $\langle p\rangle=0$ gives this minimum bound (\ref{xmin}), which is also
consistent with the above inequality (\ref{deltafrac}) and GUP relation (\ref{delta1}). Now using relation (\ref{xmin})
along with the condition $\langle p\rangle=0$ we get the maximum bound of $\Delta p$ as
\beq \Delta p_{max}=\frac{1}{\sqrt{3\beta}}\label{pmax}\eeq
It can be shown \cite{kempf} that the inequality (\ref{deltaxi}) follows from the modified Heisenberg algebra
\beq [x_{i},p_{j}]=i\hbar(\delta_{ij}+\beta\delta_{ij}p^{2}+2\beta p_{i}p_{j})~. \label{heis} \eeq
To satisfy the Jacobi identity, the above bracket (\ref{heis}) gives $[x_{i},x_{j}]=[p_{i},p_{j}]=0~$, to first order in
$O(\beta)$ \cite{kem}. Now defining
\beq x_{i}=x_{0i}~~~~~,~~~~~p_{i}=p_{0i}(1+\beta p_{0}^{2}) \label{x0i} \eeq
where $p_{0}^{2}=\sum_{j=1}^{3}~p_{0j}p_{0j}$ and $x_{0i},~p_{0j}$ satisfying the canonical
commutation relations $[x_{0i},p_{0j}]=i\hbar\delta_{ij}$, it is easy to show that the above
commutation relation (\ref{heis}) is satisfied, to first order of $\beta$. Henceforth, we neglect
terms of order $\beta^{2}$ and higher.The effects of this GUP (\ref{deltaxi}) in lamb shift and Landau levels
have been studied in \cite{das}. Also, formulation of coherent states for
this GUP has been described in \cite{ghosh}. In this work, we
successfully derive the kernel for this GUP model by Hamiltonian
path integral formulation \cite{garrod} and show that this GUP corrected kernel
induces a maximum momentum bound in the theory.\\
Using (\ref{x0i}), we start with the corresponding Hamiltonian of the form
\beq H=\frac{p^{2}}{2m}+V(\vec{r}) \label{ham} \eeq
can be written as
\beq H = H_{0} + H_{1} + O(\beta^{2}), \label{ham1} \eeq
where
\beq H_{0}=\frac{p_{0}^{2}}{2m} + V (\vec{r})~~~~,~~~~H_{1} = \frac{\beta}{m}p_{0}^{4}. \label{ham01} \eeq
Thus, we see that any system with an well defined quantum
(or even classical) Hamiltonian $H_{0}$, is perturbed
by $H_{1}$, near the Planck scale. In other words,
Quantum Gravity effects are in some sense universal.
Now the modified Schrodinger equation corresponding to the above Hamiltonian (\ref{ham1}) is
given by
\beq -\frac{\hbar^{2}}{2m}\frac{\partial^{2}}{\partial x^{2}}
\psi(x,t)+\frac{\beta\hbar^{4}}{m}\frac{\partial^{4}}
{\partial x^{4}}\psi(x,t)+V(x)\psi=i\hbar\frac{\partial}{\partial t}\psi(x,t). \label{sch} \eeq
In this paper we will see that path integral method \cite{feynman} is applicable
to this higher energy cases and we will evaluate the free particle kernel for GUP corrected
Hamiltonian (\ref{ham1}). For this purpose, we shall briefly recall the notion of basic properties
of kernel in Hamiltonian path integral formalism. The kernel in Hamiltonian path integral
is given by \cite{garrod}
\beq K(x'',x')=\int~\left[~e^{\frac{i}{\hbar} \int(\vec{p}. \vec{\dot{x}}-H)dt}\right]~\frac{dp_{1}}{2\pi\hbar}
\frac{dp_{2}}{2\pi\hbar}...\frac{dp_{N}}{2\pi\hbar}dx_{1}dx_{2}...dx_{N-1}. \label{kerhamdef} \eeq
It has been shown in \cite{garrod} that the above kernel (\ref{kerhamdef}) can be written in the form
\beq K(x'',x',\Delta t)=\delta (\vec{x}''-\vec{x}')-\frac{i\Delta t}{\hbar}
\left[-\frac{\hbar^{2}\nabla^{2}}{2m}\times\delta (\vec{x}''-
\vec{x}')+\bar{V}(x)\delta (\vec{x}''-\vec{x}')\right] \label{kerdelta} \eeq
from which one can easily obtain Schrodinger equation using the relation
\beq \psi(x'',t'')~=~\int K(x'',t'';x',t')\psi(x',t')~dx'. \label{psi} \eeq
Now taking the complex conjugate of the above equation, we have
\beq \psi^{*}(x'',t'')~=~\int~K^{*}(x'',t'';x',t')\psi^{*}(x',t')~dx'. \label{psi*} \eeq
Since $$ \int \psi^*(x'',t'') \psi(x'',t'') d x'' =
\int \psi^*(x',t') \psi(x',t') d x',$$ using the relations (\ref{psi}, \ref{psi*}) we have
\beq \int \int \int~K^{*}(x'',t'';x'_{1},t')K(x'',t'';x',t')\psi^{*}(x'_{1},t')
\psi(x',t')~dx''dx'_{1} d x' = \int \psi^{*}(x',t') \psi(x', t') dx', \label{cons} \eeq
which immediately implies the following relation
\beq \int \int~K^{*}(x'',t'';x'_{1},t')K(x'',t'';x',t')\psi^{*}(x'_{1},t')
~dx''dx'_{1}=\psi^{*}(x',t'). \label{kkf}\eeq
Also, if $\psi(x,t)$ is the solution of the Schrodinger equation
\beq -\frac{\hbar^{2}}{2m}\nabla^{2}\psi(x,t)+V(x)\psi(x,t)=i\hbar\frac{\partial \psi(x,t)}{\partial t},
\label{schf} \eeq
then the kernel also satisfy Schrodinger equation at the end point $x=x''$, i.e
\beq -\frac{\hbar^{2}}{2m}\frac{\partial^{2}}{{\partial x''}^{2}}K(x'',t'';x',t')+V(x'')K(x'',t'';x',t')=
i\hbar\frac{\partial}{\partial t''} K(x'',t'';x',t'). \label{schend} \eeq
Equation (\ref{psi}),(\ref{kkf}) and (\ref{schend}) are the basic properties of the kernel $K(x,t)$.

\section{Kernel for GUP corrected Hamiltonian:}
Path integral method \cite{feynman} is applicable in all cases where the
change of action, corresponding to the variation of path, is large enough
compared to $\hbar$. As the above Hamiltonian (\ref{ham1}) is associated
with higher energy, so a small variation on paths other than the path of
least action make enormous change in phase for which cosine or sine
will oscillate exceedingly rapidly between plus and minus value and cancel out
their total contribution. So only the least action path will contribute in kernel.
This is similar to the idea of path integral in quantum mechanics. We now
therefore consider the Hamiltonian (\ref{ham1}) in one dimension
\beq H=\frac{p_{0}^{2}}{2m}+\frac{\beta}{m}p_{0}^{4}+V(x). \label{hamfinal} \eeq
If we consider that the particle goes from $x'$ to $x''$ during the short time interval
$\Delta t$, then the kernel is of the form
\beq K(x'',t'+\Delta t~;~x',t')=\int e^{\frac{i}{\hbar}\int_{t'}^{t'+\Delta t}\left(p_{0}.\dot{x}
-\frac{p_{0}^{2}}{2m}-\frac{\beta}{m}p_{0}^{4}-V(x)\right) d t}~\frac{dp_{0}}{2\pi\hbar}
=\int e^{\frac{i}{\hbar}\left(p_{0}.(x''-x')
-\frac{p_{0}^{2}\Delta t}{2m}-\frac{\beta\Delta t}{m}p_{0}^{4}-\Delta t\bar{V}(x)\right)}
~\frac{dp_{0}}{2\pi\hbar}$$$$
=\int e^{\frac{i}{\hbar}~p_{0}.(x''-x')}e^{-\frac{i}{\hbar}
\left[\frac{p_{0}^{2}\Delta t}{2m}+\frac{\beta\Delta t}{m}p_{0}^{4}+\Delta t\bar{V}(x)\right]}
~\frac{dp_{0}}
{2\pi\hbar}, \label{kerham} \eeq
where $\bar{V}(x)$ is the average of $V(x)$ over the straight line connecting $x''$ and $x'$.
Expanding the second exponential function in (\ref{kerham}) and neglecting the
second and higher order terms of $\Delta t$, we have
\beq K(x'',t'+\Delta t~;~x',t')=\int e^{\frac{i}{\hbar}~p_{0}.
(x''-x')}\left[1-\frac{i\Delta t}{\hbar}\left(\frac{p_{0}^{2}}{2m}+\frac{\beta}{m}p_{0}^{4}
+\bar{V}(x)\right)\right]~\frac{dp_{0}}{2\pi\hbar}
$$$$=\delta(x''-x')-\frac{i\Delta t}{\hbar}\left[-\frac{\hbar^{2}}{2m}\frac{\partial^{2}}{{\partial x''}^{2}}\delta(x''-x')+\frac{\beta\hbar^{4}}{m}\frac{\partial^{4}}{{\partial x''}^{4}}
\delta(x''-x')+\bar{V}(x)\delta(x''-x')\right]. \label{kerham2} \eeq
It is interesting to note that the kernel (\ref{kerham2}) boils down to the same form
as in \cite{garrod} in the limit $\beta \longrightarrow 0$.
But it is very difficult to deal with above form of kernel (\ref{kerham2}) as it
contains derivative of delta function. Therefore we are going to derive the
delta-independent equivalent form of kernel. For this we consider the kernel
for free particle,
\beq K(x'',t'+\Delta t~;~x',t')=\int e^{\frac{i}{\hbar}\int_{t'}^{t'+\Delta t}\left(p_{0}.\dot{x}
-\frac{p_{0}^{2}}{2m}-\frac{\beta}{m}p_{0}^{4}\right) d t}~\frac{dp_{0}}{2\pi\hbar}, \label{kerfree} \eeq
for short time interval $\Delta t$. Now expanding the exponential series of
the last term in (\ref{kerfree}) and neglecting the
terms containing higher order of $\beta$ we have,
\beq K(x'',t'+\Delta t~;~x',t')=\int e^{-\frac{i \Delta t}{2 m \hbar}\left(p_{0}-\frac{(x''-x')m}{\Delta t}\right)^{2}+
\frac{i m(x''-x')^{2}}{2\hbar\Delta t}}\left(1-\frac{i ~\beta ~\Delta t ~p_{0}^{4}}{\hbar m}\right)~\frac{dp_{0}}
{2\pi\hbar}. \label{kerfree2} \eeq
After some calculation we get the kernel as,
\beq K(x'',t'+\Delta t~;~x',t')=\sqrt{\frac{m}{2\pi i\hbar\Delta t}}
\left[1+\frac{3\beta i\hbar m}{\Delta t}
-\frac{6\beta m^{2}(x''-x')^{2}}{{\Delta t}^{2}}-\frac{i\beta m^{3}(x''-x')^{4}}
{\hbar{\Delta t}^{3}}\right]e^{\frac{im(x''-x')^{2}}{2\hbar\Delta t}}. \label{kerfreeinf} \eeq
For a finite interval, we divide the time interval into N subintervals of
equal length $\Delta t$ and then we calculate the kernel as:
\beq\begin{array}{lll}
K(x'', t''; x', t') = \left(\sqrt{\frac{m}{2 \pi i \hbar \Delta t}}\right)^N \int
d x_1 d x_2 ...d x_{N-1}~~ e^{\frac{i m}{2 \hbar \Delta t}
[(x_1 - x_0)^2 + (x_2-x_1)^2+...+(x_N - x_{N-1})^2]}\\
\times \{1+\frac{3\beta i\hbar m}{\Delta t}
-\frac{6\beta m^{2}(x_{1}-x_{0})^{2}}{{\Delta t}^{2}}-\frac{i\beta m^{3}
(x_{1}-x_{0})^{4}}{\hbar{\Delta t}^{3}}\}
\{1+\frac{3\beta i\hbar m}{\Delta t}
-\frac{6\beta m^{2}(x_{2}-x_{1})^{2}}{{\Delta t}^{2}}-\frac{i\beta m^{3}(x_{2}-x_{1})^{4}}
{\hbar{\Delta t}^{3}}\}\\...
\{1+\frac{3\beta i\hbar m}{\Delta t}
-\frac{6\beta m^{2}(x_{N}-x_{N-1})^{2}}{{\Delta t}^{2}}-\frac{i\beta m^{3}(x_{N}-x_{N-1})^{4}}
{\hbar{\Delta t}^{3}}\}.
\end{array}
\label{kerfreefin} \eeq
After a bit of lengthy algebra (see appendix: 1), the final form of the kernel becomes
\beq
K(x'', t''; x', t') = \sqrt{\frac{m}{2 \pi i \hbar (t''-t')}}
e^{\frac{i m (x''-x')^2}{2 \hbar (t''-t')}}
\left[1+\frac{3\beta i\hbar m}{(t''-t')}
-\frac{6\beta m^{2}(x''-x')^{2}}{(t''-t')^{2}}-\frac{i\beta m^{3}(x''-x')^{4}}
{\hbar(t''-t')^{3}}\right],
\label{kerfreefinal} \eeq
where $t''-t' = N \Delta t$. This kernel (\ref{kerfreefinal}) is exactly of the same
form as the kernel for the infinitesimal interval (\ref{kerfreeinf}).
It can be shown that the above kernel (\ref{kerfreefinal}) satisfies the
modified schrodinger equation
\beq
-\frac{\hbar^{2}}{2m}\frac{\partial^{2}}{\partial x^{2}}\psi(x,t)
+\frac{\beta\hbar^{4}}{m}\frac{\partial^{4}}{\partial x^{4}}\psi(x,t)
=i\hbar\frac{\partial}{\partial t}\psi(x,t), \label{kersch}
\eeq
at the point $x=x'',~t=t''$ (see appendix: 2). Now the solution of
this Schrodinger equation (\ref{kersch}) is given by \cite{das}
\beq
\psi(x, t) = \left(A e^{i k(1-\beta\hbar^{2}k^{2})x-\frac{i E t}{\hbar}} +
B e^{-i k(1-\beta\hbar^{2}k^{2})x-\frac{i E t}{\hbar}}
+ C e^{\frac{x}{\sqrt{2\beta\hbar^{2}}}-\frac{i E t}{\hbar}}
+De^{-\frac{x}{\sqrt{2\beta\hbar^{2}}}-\frac{i E t}{\hbar}}\right).
\label{psif} \eeq
With this solution (\ref{psif}) in hand, we can show that the
kernels (\ref{kerham2}) and (\ref{kerfreeinf}) indeed propagates the wave function
$\psi(x, t)$ from a point $(x', t')$ to the point $(x'', t'')$,
for a chosen time interval $\Delta t=t''-t'$, such that
$\frac{\Delta t}{4\beta\hbar m}=D$=a dimensionless quantity $=o(\beta)$.
Thus the following relation holds:
\beq
\psi(x'', t'') = \int K (x'', t''; x',t') \psi(x', t') d x',
\label{prop2} \eeq
where we use the fact that
\beq
e^{\frac{-i E t'}{\hbar}}=e^{\frac{-i E t''}{\hbar}+\frac{i E\Delta t}{\hbar}}
=e^{\frac{-i E t''}{\hbar}+4i E m(\beta D)}
=e^{\frac{-i E t''}{\hbar}+4i E m.O(\beta^{2})}=e^{\frac{-i E t''}{\hbar}}
\label{ee} \eeq
Now for finite interval, we divide the time interval $(t''-t')$ into
$N$ parts of equal length $\Delta t$ in such a manner that
$\frac{\Delta t}{4\beta\hbar m}=D=o(\beta)$. Then applying equation
(\ref{prop2}) for each interval, we have
\beq
\int K(x'',t'';x',t')\psi(x',t')dx'$$$$
=\int\int...\int K(x'',t'';x_{N-1},t_{N-1})K(x_{N-1},t_{N-1};x_{N-2},t_{N-2})...K(x_{1},t_{1};x',t')
\psi(x',t')dx'dx_{1}...dx_{N-1}$$$$
=\int\int...\int K(x'',t'';x_{N-1},t_{N-1})K(x_{N-1},t_{N-1};x_{N-2},t_{N-2})...K(x_{2},t_{2};x_{1},t_{1})
\psi(x_{1},t_{1})dx_{1}dx_{2}...dx_{N-1}$$$$
=.....=\psi(x'',t'')
\label{propfinite} \eeq
Therefore for a finite interval $(t''-t')$, we have
\beq\psi(x'', t'') = \int K (x'', t''; x',t') \psi(x', t') d x'. \label{propfinal} \eeq
The alternative way to check the above relation is the condition
\beq \int\int~K^{*}(x'',t'';x'_{1},t')
K(x'',t'';x',t')\psi^{*}(x'_{1},t')~dx''dx'_{1}=\psi^{*}(x',t'). \label{com} \eeq
which is proved in appendix: 3. Therefore the free particle kernel satisfies
the basic properties of kernel, which we have stated in earlier section.\\

For usual free particle case, the probability that a particle arrives at the
point $x''$ is proportional to the absolute square of the kernel $K(x'',x',t''-t')$,
i.e. for usual free-particle kernel the probability is given by
\beq
P(x'') d x = \frac{m}{2 \pi \hbar (t''-t')} d x. \label{prob}
\eeq
Now, for the GUP corrected kernel (\ref{kerfreefinal}), the corresponding
probability is given by
\beq
P(x'') d x = K^*(x'',x',t''-t') K(x'',x',t''-t') d x =
\left(1 - \frac{12 \beta m^2 (x'' - x')^2}{(t''-t')^2}\right)
\frac{m}{2 \pi \hbar (t''-t')} d x. \label{probgup}
\eeq
It is clearly observable that the term
$ \left(1 - \frac{12 \beta m^2 (x'' - x')^2}{(t''-t')^2}\right)$
in (\ref{probgup}) is smaller than 1 as $\beta > 0$.Then we conclude that the probability value in this case is
less than the corresponding value in the free particle case. Also, since
probability is non-negative, this term $ \left(1 - \frac{12 \beta m^2 (x'' - x')^2}{(t''-t')^2}\right)$
should also be non-negative. Thus we have the following relation:
\beq
1 - \frac{12 \beta m^2 (x'' - x')^2}{(t''-t')^2} \geq 0, \label{pos}
\eeq
which immediately implies the bound for momentum as
\beq
p \leq p_{max} = \frac{1}{2 \sqrt{3 \beta}}; \label{pmax1}
\eeq
where $p = \frac{m x}{t}$. Thus, from (\ref{pmax1}) we see that GUP induces a momentum upper
bound in the theory which is comparable to maximum momentum uncertainty (\ref{pmax}) induced by GUP.\\
If we consider higher order terms of $\beta$, i.e. terms up to $o(\beta^2)$, the Hamiltonian (\ref{ham}) becomes
\beq H=\frac{p_0^2}{2m}+\frac{\beta p_0^4}{m}+\frac{\beta^2 p_0^6}{2m}\label{betahighham}\eeq
and the corresponding path integral becomes
\beq K(x'',t'+\Delta t~;~x',t')=\int e^{\frac{i}{\hbar}[p_0.\dot{x}-\frac{p_0^2}{2m}-\frac{\beta p_0^4}{m}-\frac{\beta^2 p_0^6}{2m}]}\frac{dp_0}{2\pi\hbar}.\label{betahighker}\eeq
Proceeding in the same way as before we get the final form of kernel as
\beq K(x'',t'+\Delta t~;~x',t')=\sqrt{\frac{m}{2\pi i\hbar\Delta t}}
\left[\left(1-\frac{6\beta m^{2}(x''-x')^{2}}{{\Delta t}^{2}}-\frac{105\beta^2\hbar^2 m^2}{2\Delta t^2}\right.\right.$$$$
\left.\left.+\frac{105\beta^2\hbar^2 m^4(x''-x')^4}{\Delta t^4}-\frac{\beta^2m^6 (x''-x')^8}{2\hbar^2\Delta t^6}+\frac{15\beta^2\hbar^2 m^2}{2\Delta t^2}-\frac{15\beta^2 m^4 (x''-x')^4}{2\Delta t^4}\right)\right.$$$$
\left.+i\left(\frac{3\beta\hbar m}{\Delta t}-\frac{\beta m^{3}(x''-x')^{4}}{\hbar{\Delta t}^{3}}-\frac{210\beta^2 \hbar m^3 (x''-x')^2}{\Delta t^3}+\frac{14\beta^2 m^5 (x''-x')^6}{\hbar\Delta t^5}\right.\right.$$$$
\left.\left.+\frac{45\beta^2 \hbar m^3 (x''-x')^2}{2\Delta t^3}-\frac{\beta^2 m^5 (x''-x')^6}{2\hbar\Delta t^5}\right)\right]
e^{\frac{im(x''-x')^{2}}{2\hbar\Delta t}}\eeq
Now, as we know that the final form of kernel is same for infinitesimal and finite interval, so we can deal with the above form of kernel for infinitesimal
interval as well. In this case, the probability of finding the particle at the region $dx$ enclosing the point $x''$ is
\beq K^*(x'',t'';x',t')K(x'',t'';x',t')~dx=\frac{m}{2\pi\hbar\Delta t}\left(1-\frac{12\beta m^2 (x''-x')^2}{\Delta t^2}-\frac{81\beta^2 \hbar^2 m^2}{\Delta t^2}
+\frac{225\beta^2 m^4 (x''-x')^4}{\Delta t^4}\right)~dx$$$$
=\frac{m}{2\pi\hbar\Delta t}\left(1-12\beta p^2+225\beta^2 p^4-\frac{81\beta^2 \hbar^2 m^2}{\Delta t^2}\right)~dx,\label{betahighprob}\eeq
where we have neglected the terms of higher order than $\beta^2$. Since probability is always non-negative, we have the following relation
\beq 1-12\beta p^2+225\beta^2 p^4-\frac{81\beta^2 \hbar^2 m^2}{\Delta t^2}\geqslant 0\eeq
from which we obtain the bound
\beq p_{max}=\frac{1}{2\sqrt{3\beta}}\left(1-\frac{1}{2}\left(\frac{9\beta\hbar m}{\Delta t}\right)^2\right)$$$$
\approx \frac{1}{2\sqrt{3 \beta}}e^{-\frac{1}{2}(\frac{9 \beta \hbar m}{\Delta t})^2}. \label{bpbound} \eeq
From (\ref{bpbound}) we clearly see that the maximum momentum bound for $o(\beta^2)$ case is less than that obtained in
the previous $o(\beta)$ case, though by a very small amount with the correction terms being of the order of or higher
than $o(\beta^2)$ .
\section{Conclusions and future prospects:}
We know that GUP gives rise additional terms in quantum mechanical
Hamiltonian like $\beta p^{4}$, where $\beta\sim\frac{1}{(M_{pl}c)^{2}}$
is the GUP parameter. This term plays important role at Planck energy level.
Considering this term as a perturbation, we have shown that path integral is
applicable on this GUP corrected non-relativistic cases. Here we have constructed
the two form of kernel by applying Hamiltonian path integral method. The consistency
properties of this kernel is then thoroughly verified. We have shown that the
probability for finding a particle at a given point in case of the GUP model
is less than the corresponding probability in the usual free particle case. We
have also shown that probabilistic interpretation of this kernel induces a
momentum upper bound in the theory. And this upper bounds changes slightly
with $e^{-o(\beta^2)}$, if we consider higher order term of
$\beta$ in the Hamiltonian (\ref{ham}). Now following the Hamiltonian path
integral approach one can construct kernels and study their properties for
other systems like particle in a step potential, Hydrogen atom etc, which is
our future goal.
\section{Acknowledgements:}
We would like to thank Prof. Subir Ghosh, for suggesting the problem and discussion with us.

\section{Appendix: 1}
Neglecting the term containing higher order of $\beta$ equation (\ref{kerfreefin}), becomes
\beq K(x'',t'';x',t')=\sqrt{\frac{m}{2\pi i\hbar \Delta t}}\int
~~ e^{\frac{i m}{2 \hbar \Delta t}[(x_1-x_0)^2 + (x_2-x_1)^2+...+(x_N - x_{N-1})^2]}
\left[1+\frac{3\beta i\hbar m}{\Delta t}N-\frac{6\beta m^{2}}{{\Delta t}^{2}}\{(x_{1}-x_{0})^{2}+\right.$$$$\left.(x_{2}-x_{1})^{2}+...+(x_{N}-x_{N-1})^{2}\}
-\frac{i\beta m^{3}}{\hbar{\Delta t}^{3}}\{(x_{1}-x_{0})^{4}+(x_{2}-x_{1})^{4}+
...+(x_{N}-x_{N-1})^{4}\}\right]d x_1 d x_2 ...d x_{N-1}~.\eeq
With the substitution  $x_i=\sqrt{\frac{2\hbar\Delta t}{m}}y_i$  the above equation becomes
\beq K(x'',t'';x',t')=\sqrt{\frac{m}{2\pi i\hbar \Delta t}}\left(\frac{1}{i\pi}\right)^{\frac{N-1}{2}}\int
~~ e^{i(y_1-y_0)^2+i(y_2-y_1)^2+...+i(y_N - y_{N-1})^2}
\left[1+\frac{3\beta i\hbar m}{\Delta t}N-\frac{12\beta\hbar m}{\Delta t}\{(y_{1}-y_{0})^{2}+\right.$$$$\left.(y_{2}-y_{1})^{2}+...+(y_{N}-y_{N-1})^{2}\}
-\frac{4i\beta\hbar m}{\Delta t}\{(y_{1}-y_{0})^{4}+(y_{2}-y_{1})^{4}+
...+(y_{N}-y_{N-1})^{4}\}\right]d y_1 d y_2 ...d y_{N-1}~.\eeq
Now
\beq\int e^{i(y_1-y_0)^2+...+i(y_N - y_{N-1})^2}\{(y_{1}-y_{0})^{2}+...+(y_{N}-y_{N-1})^{2}\}dy_1dy_2...dy_{N-1}$$$$
=\left\{\frac{i(N-1)}{2}+\frac{(y_N-y_0)^{2}}{N}\right\}\frac{(i\pi)^{\frac{N-1}{2}}}{\sqrt{N}}e^{\frac{i(y_N-y_0)^2}{N}}\eeq
which is done by adding $N$ separate integral of the form
\beq\int e^{i(y_1-y_0)^2+...+i(y_N - y_{N-1})^2}(y_{i}-y_{i-1})^{2}dy_1dy_2...dy_{N-1}\eeq
Again
\beq\int e^{i(y_1-y_0)^2+...+i(y_N - y_{N-1})^2}\{(y_{1}-y_{0})^{4}+...+(y_{N}-y_{N-1})^{4}\}dy_1dy_2...dy_{N-1}$$$$
=\left[-\frac{3}{4}\left\{(\frac{1^{2}}{2^{2}}+\frac{2^{2}}{3^{2}}+...+\frac{{N-1}^{2}}{N^{2}})+(
\frac{1}{1.2^{2}}+\frac{1}{2.3^{2}}+...+\frac{1}{(N-1).N^{2}})
+2(\frac{1}{2}+\frac{1}{3}+...+\frac{1}{N})-\frac{2(N-1)}{N}\right\}\right.$$$$\left.+\frac{3i(N-1)}{N}(1+1+...+1)_{N times}\frac{(y_N-y_0)^{2}}{N^{2}}+\frac{(y_N-y_0)^{4}}{N^{4}}N\right]
\frac{(i\pi)^{\frac{N-1}{2}}}{\sqrt{N}}e^{\frac{i(y_N-y_0)^2}{N}}$$$$
=\left[-\frac{3(N-1)^{2}}{4N}+3i(N-1)\frac{(y_N-y_0)^{2}}{N^{2}}+\frac{(y_N-y_0)^{4}}{N^{3}}\right]
\frac{(i\pi)^{\frac{N-1}{2}}}{\sqrt{N}}e^{\frac{i(y_N-y_0)^2}{N}}
\eeq
Therefore
\beq K(x'',t'';x',t')=\sqrt{\frac{m}{2\pi i\hbar \Delta t}}(\frac{1}{i\pi})^{\frac{N-1}{2}}\frac{(i\pi)^{\frac{N-1}{2}}}{\sqrt{N}}\left[1+\frac{3i\beta\hbar m}{\Delta t}(N-2N+2+\frac{(N-1)^{2}}{N})\right.$$$$\left.-\frac{12\beta\hbar m}{\Delta t}(N-N+1)\frac{(y_N-y_0)^{2}}{N^{2}}-\frac{4i\beta\hbar m}{\Delta t}\frac{(y_N-y_0)^{4}}{N^{3}}\right]$$$$
=\sqrt{\frac{m}{2\pi i\hbar N\Delta t}}\left[1+\frac{3i\beta\hbar m}{N\Delta t}-\frac{12\beta\hbar m}{\Delta t}\frac{(y_N-y_0)^{2}}{N^{2}}-\frac{4i\beta\hbar m}{\Delta t}\frac{(y_N-y_0)^{4}}{N^{3}}\right]e^{\frac{i(y_N-y_0)^2}{N}}.\eeq
Replacing $y_i$ by $x_i$, and kipping in mind $N\Delta t=t''-t'$, the above equation becomes (\ref{kerfreefinal}).

\section{Appendix:2}
Now differentiating $K(x'',t'+\Delta t;x',t')$ in (\ref{kerfreefinal}) partially w.r.t.
$t''$ where $\Delta t=t''-t'$ we have
\beq \frac{\partial}{\partial t''}K(x'',t'';x',t')
=e^{\frac{im(x''-x')^{2}}{2\hbar\Delta t}}\sqrt{\frac{m}{2\pi i\hbar\Delta t}}
\left(-\frac{1}{2\Delta t}-\frac{9i\beta\hbar m}{2{\Delta t}^{2}}
-\frac{i m x^{2}}{2\hbar{\Delta t}^{2}}+\frac{33\beta m^{2}x^{2}}
{2{\Delta t}^{3}}+\frac{13 i\beta m^{3}x^{4}}{2\hbar{\Delta t}^{4}}
-\frac{\beta m^{4}x^{6}}{2\hbar^{2}{\Delta t}^{5}}\right)\label{dif1t}\eeq
Similarly differentiating $K(x'',t'+\Delta t;x',t')$
partially w.r.t. $x''$ two and four times we have
\beq \frac{\partial^{2}}{{\partial x''}^{2}}K(x'',t'';x',t')=e^{\frac{im(x''-x')^{2}}{2\hbar\Delta t}}\sqrt{\frac{m}
{2\pi i\hbar\Delta t}}
\left(\frac{i m}{\hbar\Delta t}
-\frac{15\beta m^{2}}{{\Delta t}^{2}}-\frac{45\beta i m^{3}x^{2}}
{\hbar{\Delta t}^{3}}+\frac{15\beta m^{4}x^{4}}{\hbar^{2}{\Delta t}^{4}}-\frac{m^{2}x^{2}}{\hbar^{2}{\Delta t}^{2}}
+\frac{i\beta m^{5}x^{6}}{\hbar^{3}{\Delta t}^{5}}\right)\label{dif2x}\eeq
and
\beq \frac{\partial^{4}}{{\partial x''}^{4}}K(x'',t'';x',t')
=e^{\frac{im(x''-x')^{2}}{2\hbar\Delta t}}\sqrt{\frac{m}{2\pi i\hbar\Delta t}}
\left(-\frac{6i m^{3}x^{2}}{\hbar^{3}{\Delta t}^{3}}+\frac{m^{4}x^{4}}{\hbar^{4}{\Delta t}^{4}}
-\frac{3m^{2}}{\hbar^{2}{\Delta t}^{2}}\right)\label{dif4x}\eeq
Now multiply (\ref{dif2x}) by $\frac{-\hbar^{2}}{2m}$ and (\ref{dif4x}) by $\frac{\beta\hbar^{4}}{m}$ and
after adding this two we see that this is $i\hbar$ times of (\ref{dif1t}), which generates the equation
\beq i\hbar\frac{\partial}{\partial t''}K(x'',t'';x',t')
=\frac{-\hbar^{2}}{2m}\frac{\partial^{2}}
{{\partial t''}^{2}}K(x'',t'';x',t')+\frac{\beta\hbar^{4}}{m}
\frac{\partial^{4}}{{\partial t''}^{4}}K(x'',t'';x',t')\eeq
where $\Delta t=t''-t'$. Which is modified schrodinger equation for free particle.
\section{Appendix: 3}
Taking the complex conjugate of the equation (\ref{kerfreefinal}), we have
\beq
K^{*}(x'', t''; x', t') = \sqrt{\frac{-m}{2 \pi i \hbar (t''-t')}}
e^{\frac{-i m (x''-x')^2}{2 \hbar (t''-t')}}
\left[1-\frac{3\beta i\hbar m}{(t''-t')}
-\frac{6\beta m^{2}(x''-x')^{2}}{(t''-t')^{2}}+\frac{i\beta m^{3}(x''-x')^{4}}
{\hbar(t''-t')^{3}}\right]~.
\eeq
Therefore
\beq\int K^{*}(x'', t''; x_{1}', t')K(x'', t''; x', t')dx''$$$$
=\left[\left(1-\frac{6\beta m^{2}(x'-x_{1}')^{2}}{(t''-t')^{2}}
-\frac{6i\beta m^{3}(x'-x_{1}')^{4}}{\hbar(t''-t')^{3}}\right)\delta(x'-x_{1}')-\left(\frac{12i\beta m\hbar(x'-x_{1}')}{(t''-t')}+\frac{4\beta m^{2}(x'-x_{1}')^{3}}{\hbar(t''-t')^{2}}\right)\delta'(x'-x_{1}')\right.$$$$\left.
+\left(12\beta\hbar^{2}-\frac{6i\beta\hbar m(x'-x_{1}')^{2}}{(t''-t')}\right)\delta''(x'-x_{1}')
+4\beta\hbar^{2}(x'-x_{1}')\delta'''(x'-x_{1}')\right]e^{\frac{-i m (x'-x_{1}')^2}{2 \hbar (t''-t')}},\label{k*kdx''}\eeq
where we have used the relation
\beq\int e^{-u.(x'-x'_1)}u^{n}du~=2\pi(i)^{n}\frac{d^{n}\delta(x'-x'_1)}{d(x'-x'_1)^{n}}~.\eeq
Now multiplying both side  of (\ref{k*kdx''}) by $\psi^{*}(x_{1}',t')$ and integrating w.r.to $x_{1}'$  we have
\beq\int\int~K^{*}(x'',t'';x'_{1},t')K(x'',t'';x',t')\psi^{*}(x'_{1},t')~dx''dx'_{1}$$$$
=\int\left[\left(1-\frac{6\beta m^{2}(x'-x_{1}')^{2}}{(t''-t')^{2}}
-\frac{6i\beta m^{3}(x'-x_{1}')^{4}}{\hbar(t''-t')^{3}}\right)\delta(x'-x_{1}')\psi^{*}(x'_1,t')\right.$$$$
\left.-\left(\frac{12i\beta m\hbar(x'-x_{1}')}{(t''-t')}+\frac{4\beta m^{2}(x'-x_{1}')^{3}}
{\hbar(t''-t')^{2}}\right)\delta'(x'-x_{1}')\psi^{*}(x'_1,t')
\right.$$$$\left.+\left(12\beta\hbar^{2}-\frac{6i\beta\hbar m(x'-x_{1}')^{2}}{(t''-t')}\right)\delta''(x'-x_{1}')\psi^{*}(x'_1,t')
+4\beta\hbar^{2}(x'-x_{1}')\delta'''(x'-x_{1}')\psi^{*}(x'_1,t')\right]
e^{\frac{-i m (x'-x_{1}')^2}{2 \hbar(t''-t')}}dx'_1.\eeq
Using the relation
\beq \int f(x)\delta^{n}(x)dx=-\int\frac{\partial f(x)}{\partial x}\delta^{n-1}(x)dx\eeq
we get from above
\beq\int\int~K^{*}(x'',t'';x'_{1},t')K(x'',t'';x',t')\psi^{*}(x'_{1},t')~dx''dx'_{1}=\psi^{*}(x',t')~.\eeq

\ed